\documentclass[11pt,a4paper]{article}

\usepackage[margin=1in]{geometry}
\usepackage{amsmath}
\usepackage{amssymb}
\usepackage{graphicx}
\usepackage{bm}
\usepackage{mathtools}
\usepackage{authblk}
\usepackage{cite}
\usepackage{color}
\usepackage{hyperref}
\usepackage{multibib}
\usepackage{comment}

\title{\Large Optical Skyrmions in Waveguides}
\date{}
\author[1,*]{An Aloysius Wang}
\author[1]{Yifei Ma}
\author[1]{Yuxi Cai}
\author[1]{Ji Qin}
\author[2]{Bowei Dong}
\author[1,*]{Chao He}
\affil[1]{\small Department of Engineering Science, University of Oxford, Parks Road, Oxford, OX1 3PJ, UK}
\affil[2]{\small Institute of Microelectronics (IME), Agency for Science, Technology and Research (A\text{*}STAR), 2 Fusionopolis Way, Innovis \#08-02, Singapore 138634, Republic of Singapore}
\affil[*]{Corresponding authors: aw6609@princeton.edu; chao.he@eng.ox.ac.uk}

\renewcommand{\figurename}{Fig.}

\begin{document}
\maketitle

\renewcommand{\figurename}
{Fig.}

{\bf Optical skyrmions are topologically non-trivial polarization fields which have recently attracted attention due to their potential use in high density data applications such as optical communications, photonic computing and more. An important hurdle in utilizing optical skyrmions for such applications is establishing conditions under which their topological structure remains preserved during propagation: while results of this type already exist for paraxial beams in free-space propagation, the critical case relevant to modern applications involves propagation in confined media, such as waveguide systems. In this paper, we demonstrate for the first time that, within a conducting waveguide, the preservation of the skyrmion number during propagation is determined by the presence of so-called topologically stabilizing modes. If such a mode is present, not only will topological protection hold despite the transverse polarization profile changing due to modal dispersion, but there is also a degree of robustness to variations in the coefficients of TE and TM modes present. Lastly, we demonstrate how a generalized skyrmion number can recover topological protection in situations where the usual skyrmion number is not preserved. Our methods open new avenues for robust high-dimensional information manipulation in waveguiding systems.} \\

Skyrmions are a category of topological solitons which were first discovered by the British physicist Tony Skyrme as a candidate model of the nucleon \cite{skyrme1961non}. Since then, skyrmions have been considered in a huge number of different domains, including magnetic materials \cite{Fert2017,Nagaosa2013}, Bose-Einstein condensates \cite{al_khawaja_skyrmions_2001}, liquid crystals \cite{foster_two-dimensional_2019}, and more \cite{PhysRevLett.132.054003}. Most recently, skyrmions have been observed in optics and photonics \cite{Tsesses2018, He2022, Shen2022, Bai2020, du_deep-subwavelength_2019, lei_photonic_2021, shi_spin_2021, Gao2020, wang2024unlock, teng_physical_2023, Lin2021, shi_strong_2020, he2023universal}, and are hypothesized to have potential applications in areas such as imaging, quantum technologies, information storage, and more \cite{shen_optical_2023, ornelas_non-local_2024}. 

In this paper, we focus on one particularly promising use of the optical stokes skyrmion, namely high-density data applications such as on-chip optical communications and photonic computing \cite{wang2024unlock}. The motivation behind the use of optical skyrmions in such settings is threefold. Firstly, skyrmions tap into the vectorial properties of light, representing a step towards higher-dimensional information. Secondly, the naturally discrete nature of the skyrmion number interfaces perfectly with digital information. Thirdly, the topological nature of the skyrmion suggests a robustness to perturbations which is essential for reliable information manipulation and transfer. However, in order to utilize optical skyrmions in real applications, it is necessary to first establish their topological preservation during propagation. While this has been demonstrated in free space under restricted conditions, such as in beams formed via a superposition of Laguerre-Gaussian (LG) modes, it is not true in general \cite{Liu2022}. As the topological nature of the skyrmion depends primarily on its boundary conditions \cite{Complex}, confined media provide a natural setting for exploring topological behavior. In particular, confined media have well-defined boundaries that not only simplify the analysis considerably but also, as we show in this paper, play a crucial role in enabling topological protection.

From the perspective of modern applications, waveguides play a central role in many important photonic systems and are crucial in enabling high-bandwidth, low-latency manipulation of data. This makes establishing topological protection in waveguides a significant task, as an increase in data transfer along waveguides has ramifications for many different and important technologies including photonic integrated circuits, biosensing, quantum computing, artificial intelligence, and more \cite{waveguide, Marshall:09}.

In real-world implementations, single-mode waveguides are used to avoid modal dispersion. However, this approach sacrifices efficiency due to the difficulty of coupling light into such small waveguides and forgoes the opportunity to exploit multiple modes to enhance information transfer. Nonetheless, the efficient use of multi-mode waveguides remains a difficult engineering challenge \cite{LiLiuDai+2019+227+247}, with modal dispersion a central obstacle to maintaining signal integrity. To address this, we introduce the notion of \textit{topologically stabilizing}, \textit{weakly stable} and \textit{unstable} modes, and apply this classification to establish conditions under which the skyrmion number integral within a multi-mode waveguide is preserved during propagation, even in the presence of dispersion. This provides a potential method to robustly exploit the capacity to support multiple modes in larger waveguides for information transfer and manipulation, which could significantly increase bandwidth and enable more complex on-chip photonic functionalities.

Note that the synthesis of optical skyrmions via confined media, such as through metafibers \cite{he2024metafiber} or gradient-index systems \cite{Chao2019, Shen2023} has been observed before. Our work differs from these previous results by demonstrating a completely new method of manipulating and transporting topological information through waveguides that is both robust and resistant to modal dispersion.

\section{Main}

As explained in \cite{Complex}, the topological nature of a traditional skyrmion arises from constraints on its boundary values. In particular, if the polarization state of an optical skyrmion approaches a constant value on its boundary, one may ``compactify'' its domain into a sphere and classify its topology by the degree of the compactified function. In the context of a conducting waveguide, this argument breaks down and it is no longer true that the polarization field within the waveguide can necessarily be compactified. However, the geometry of the waveguide entirely determines the polarization state at the air-conductor interface, and this restriction is sufficient in recovering non-trivial topologies \cite{wang2024generalizedskyrmions}. Heuristically, one can understand the statement above in the following way. Consider a conducting waveguide with a smooth boundary and a constant cross-section along its length. The polarization state along the boundary of the waveguide is then fixed with the same profile on each cross-section. Next, construct any smooth $S^2$-valued function defined on the outside of the waveguide with the following properties (1) it is a continuous extension of the field within the waveguide and (2) it approaches a constant value at infinity. Then, if the polarization state within the waveguide is everywhere well-defined, the field can be canonically extended to a continuous and compactifiable function, and can therefore be assigned a topological number through this extension. The skyrmion number integral evaluated within the waveguide is thus quantized in integer steps, with each level corresponding to a specific homotopy class of the extended function.

Since the skyrmion number integral evaluated within the waveguide is quantized, one should naturally expect topological protection to hold in propagation. Indeed, if the polarization state within the waveguide is everywhere well-defined, propagation along the waveguide is a homotopy of the field, and therefore the topological charge within the waveguide remains constant even if the transverse polarization profile changes due to effects such as dispersion. However, an important subtlety to this argument is the fact that polarization singularities (points where the electric field vanishes) can certainly occur, and this qualitatively changes the topology of the field by puncturing a hole into its domain. Therefore, a simple and sufficient condition for topological protection is to determine when singularities can be ruled out. 

One way to do this is to consider solutions to the Helmholtz equation within the waveguide whose transverse electric field strength is everywhere non-zero. Since the electric field distribution is smooth due to ellipticity of the Helmholtz equation \cite{evans}, and the map from electric field to polarization state is smooth everywhere away from zero, being everywhere non-zero guarantees that the polarization state is well-defined and smooth everywhere within the waveguide. Moreover, if this positivity condition is met by a particular solution, then clearly it is also met by any other solution sufficiently close. In the usual setting where we expand the field distribution within a waveguide as a superposition of different TE and TM modes, our discussion above can be made precise as follows: Suppose we write the electric field within the waveguide as
\begin{equation}
    \mathbf{E}_0 = \sum \alpha_{mn}\mathbf{E}^{\text{TE}_{mn}} + \beta_{mn} \mathbf{E}^{\text{TM}_{mn}}
\end{equation}
where $\alpha_{mn}$, $\beta_{mn}$ are complex coefficients. If the collection of coefficients $(\alpha_{mn},\beta_{mn})$ is such that $\mathbf{E}_0$ is everywhere non-zero, then it is easy to see that the following statements hold.

\begin{enumerate}
    \item The skyrmion number integral remains constant along the length of the waveguide, indicating topological protection during propagation. In particular, the field’s topological information can be reliably maintained and transported, even in the presence of modal dispersion.
    
    \item If $(\tilde\alpha_{mn}, \tilde\beta_{mn})$ is another set of coefficients such that 

    \begin{equation}
        \mathbf{E}_t = \sum (t\tilde\alpha_{mn} + (1-t)\alpha_{mn})\mathbf{E}^{\text{TE}_{mn}} + (t\tilde\beta_{mn} +(1-t){\beta}_{mn})\mathbf{E}^{\text{TM}_{mn}}
    \end{equation}

    is non-zero for every $t \in [0,1]$, then the skyrmion number of the field corresponding to the coefficients $(\tilde\alpha_{mn},\tilde\beta_{mn})$ is the same as the skyrmion number of the field corresponding to $(\alpha_{mn},\beta_{mn})$. Moreover, this is guaranteed to be true for some range of coefficients about $(\alpha_{mn},\beta_{mn})$. Thus there is topological robustness to parameters. From a practical standpoint, this tolerance to coefficient variations implies that the precision of coupling can be relaxed while still ensuring that the correct topological information is encoded within the field. 
    
    \item Lastly, if $(\gamma_{mn},\delta_{mn})$ is a solution which is not everywhere non-zero, then ($\gamma_{mn}+c\alpha_{mn}, \delta_{mn} + c\beta_{mn})$ is everywhere non-zero for $\lvert c \rvert$ sufficiently large.
\end{enumerate}
In light of the third point, we call any such collection of coefficients a \textit{topologically stabilizing mode} (even though the solution itself need not be a mode in the conventional sense) as adding such a mode to a state that is not topologically protected allows one to recover topological protection.

Note that if $(\alpha_{mn},\beta_{mn})$ are such that $\mathbf{E}_0$ is not everywhere non-zero, this does not immediately preclude topological protection in propagation. In particular, if there is only one mode propagating within the waveguide, then the polarization profile is constant along the length of the waveguide, and thus, so is the skyrmion number regardless of whether or not the mode possesses a singularity. However, in this scenario, there is no longer tolerance to perturbations in coefficients as a small change in the field can drastically change the polarization profile near the singularity. Such solutions are therefore \textit{weakly topologically protected} in the sense that they are robust in propagation but not to parameters.  

To make our arguments above more concrete, we specialize to the situation of a cylindrical conducting waveguide. Here, we consider a waveguide of radius $a$ extending in the $z$ direction. The general solution within the waveguide can then be written as a linear combination of TE and TM modes
\begin{equation}
    \bm{E}_0(r,\theta,z) = \sum \bm{E}^{\text{TE}_{mn}}(r,\theta)e^{ik_nz} +\sum \bm{E}^{\text{TM}_{mn}}(r,\theta)e^{i\kappa_n z},
\end{equation}
where the full expressions of individual modes are given in Methods 1. Note that we have subsumed the arbitrary coefficients $A_{mn}$, $B_{mn}$, $C_{mn}$ and $D_{mn}$ within $\bm{E}^{\text{TE}_{mn}}$ and $\bm{E}^{\text{TM}_{mn}}$. As established earlier, for the skyrmion number to be preserved along $z$, we need only check that the magnitude of the transverse component is nowhere zero. It is, however, difficult to come up with general conditions on the coefficients that guarantee this to be true. Instead, we make the following observation. Note that for both the TE and TM modes, one of $E_r$ and $E_\theta$ varies as $mJ_m(cr)/r$ while the other varies as $J_m'(cr)$ for some constant $c$, where $J_m$ is the order $m$ Bessel function of the first kind. Whenever $m \neq 1$, both of these terms evaluate to $0$ at $r=0$. Thus, among pure TE and TM modes, only $\text{TE}_{1n}$ and $\text{TM}_{1n}$ can be topologically stabilizing.

We further demonstrate some subtle features of the theory by numerical simulations. In the following, we consider TE modes within a waveguide of radius $a=50$\textmu m, light of wavelength $\lambda = 500$nm, and use a Gauss quadrature rule for all numerical integration. 

\begin{figure}[!ht]
\centering
\includegraphics[width=0.95\textwidth]{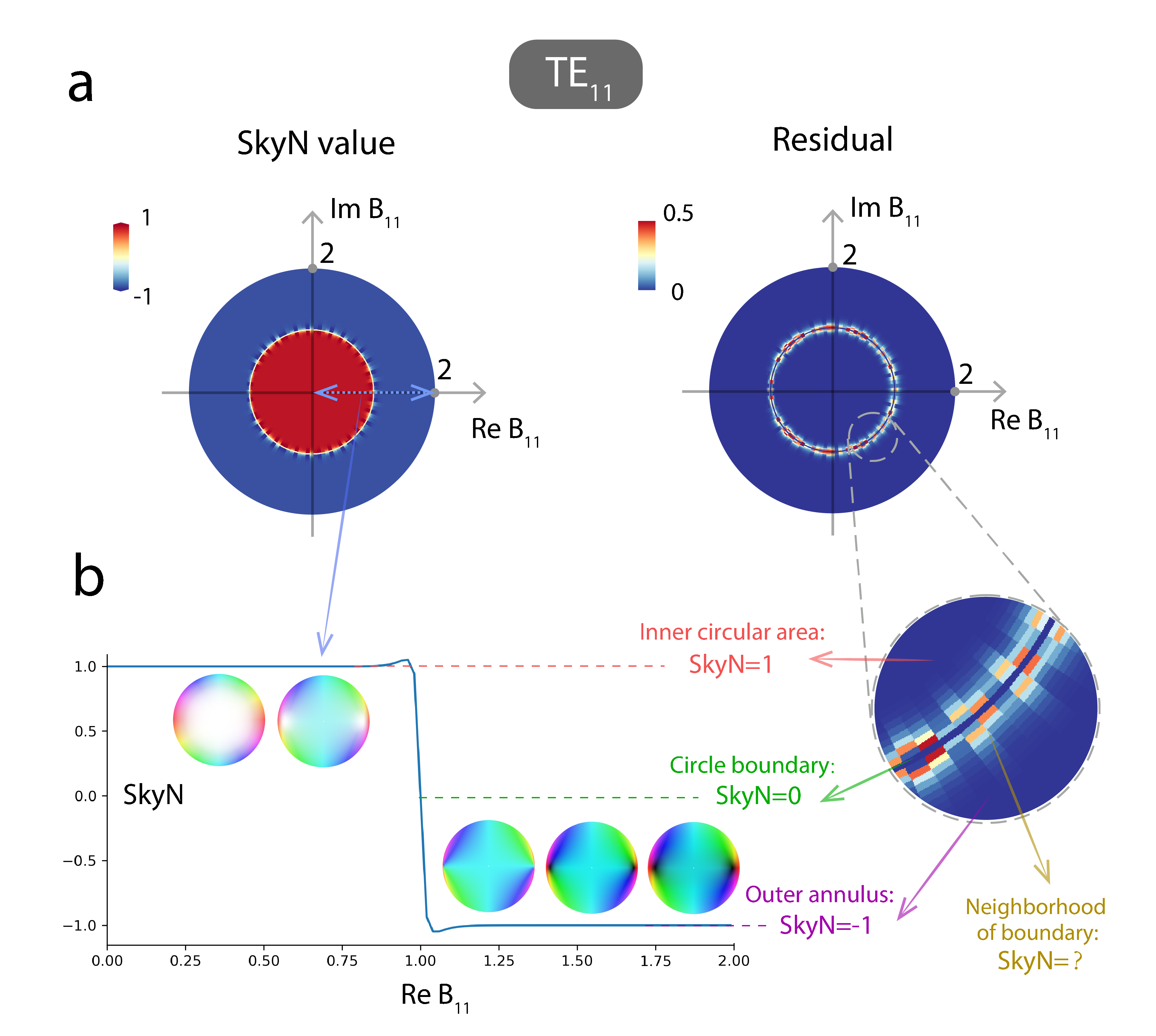}
\caption{\footnotesize \label{fig: TE11} {\bf Phase diagram of TE\textsubscript{11} mode.} a, (Left) Skyrmion number at different values of $B_{11}$. (Right) Absolute difference between computed skyrmion number and its nearest integer. The skyrmion number is stable with the value of $1$ in the region $\lvert B_{11}\rvert \lesssim 1$ and stable with a value of $-1$ in the region $\lvert B_{11}\rvert \gtrsim 1$. Between the two stable regions, there is a critical region where the behaviour of the skyrmion number is unstable, with the exception of $\lvert B_{11} \rvert = 1$ where it is exactly 0. b, Skyrmion number along the $\text{Im}\, B_{11}=0$ line. The stokes fields at different $\text{Re}\, B_{11}$ taken uniformly between 0 and 2 are also shown. }
\end{figure}

First, we study the dispersionless case: we take a single $\text{TE}$ mode (see Methods 1, equations \ref{eq: TE1}-\ref{eq: TE2}), and vary the coefficients $A_{mn}$ and $B_{mn}$ to produce different fields. Figure \ref{fig: TE11}a shows the case of $\text{TE}_{11}$ where $A_{11}=1$ is fixed and $B_{11}$ is varied. On the left, we have the numerically evaluated skyrmion number integral for different values of $B_{11}$ and on the right is the difference between the computed skyrmion number and its closest integer. Figure \ref{fig: TE11}b shows the variation of the skyrmion number along the $\text{Im}\,B_{11} = 0$ line. 

We note the following important features: firstly, as mentioned above, the $\text{TE}_{11}$ mode does not have a singularity at $r=0$, and is therefore a candidate for a topologically stabilizing mode. Consequently, we expect the skyrmion number of the mode to be robust to the coefficients $A_{11}$ and $B_{11}$. From the figure, we see that this is indeed true for most values of $B_{11}$, with the skyrmion number stable with the value of $1$ in the region $\lvert B_{11}\rvert  \lesssim 1$ and stable with the value of $-1$ in the region $\lvert B_{11}\rvert \gtrsim 1$. There is, however, a critical transition that takes place near $\lvert B_{11}\rvert = 1$ as the skyrmion number transitions from $1$ to $-1$. In this critical region the behaviour of the skyrmion number is unpredictable and it is clear that small perturbations to $B_{11}$ can have drastic impact on the skyrmion number. Mathematically, the existence of this critical region can be understood from the fact that for values $\lvert B_{11}\rvert \approx 1$, the mode fails to be everywhere non-zero.

Secondly, there is an additional phenomenon that takes place when $\lvert B_{11}\rvert=1$ exactly. In this case, the polarization state of the field is everywhere linearly polarized, and as a result, has a skyrmion number of exactly 0. This represents an intermediate topological phase which is stable to perturbations in coefficients provided the additional constraint of $\lvert B_{11}\rvert =1$ is satisfied.  

\begin{figure}[!ht]
\centering
\includegraphics[width=0.95\textwidth]{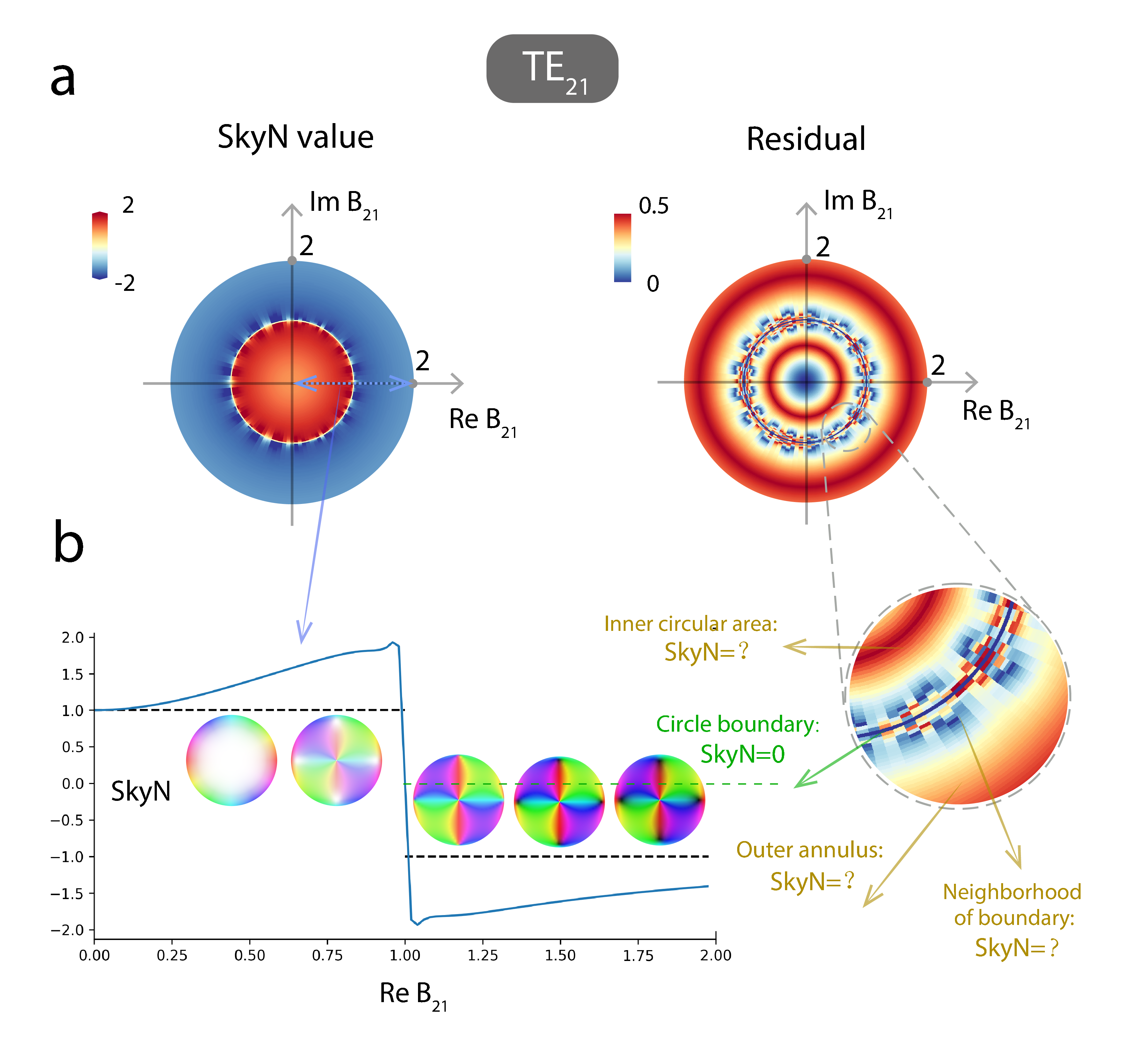}
\caption{\footnotesize \label{fig: TE21} {\bf Phase diagram of TE\textsubscript{21} mode.} a, (Left) Skyrmion number at different values of $B_{21}$. (Right) Absolute difference between computed skyrmion number and its nearest integer. The phase diagram exhibits no stable regions, but is exactly $0$ along the circle given by $\lvert B_{21}\rvert =1$. b, Skyrmion number along the $\text{Im}\, B_{11}=0$ line. Note that the skyrmion number is exactly 1 when $B_{21}=0$, and approaches -1 when $\lvert B_{21}\rvert$ approaches infinity. The stokes fields at different $\text{Re}\, B_{11}$ taken uniformly between 0 and 2 are also shown.}
\end{figure}

Figure \ref{fig: TE21} shows the case of $\text{TE}_{21}$ where $A_{21}=1$ is fixed and $B_{21}$ is varied. As above, the numerically evaluated skyrmion number integral for different values of $B_{21}$ is shown on the left, and the difference between the computed skyrmion number and its closest integer on the right. Figure \ref{fig: TE21}b shows the variation of the skyrmion number along the $\text{Im}\, B_{21}=0$ line. 

Since $\text{TE}_{21}$ has a singularity at $r=0$, we do not expect the skyrmion number to be stable with respect to coefficients. And indeed, this is clear from the figure, where the fractional part of the skyrmion number is seen to vary with $B_{21}$. There are, however, still interesting distinguished points. Firstly, as in the case of $\text{TE}_{11}$, when $\lvert B_{21}\rvert = 1$ the field is everywhere linearly polarized and thus has a skyrmion number of $0$. Additionally, the points $B_{21} = 0$ and $\lvert B_{21} \rvert \rightarrow \infty$ are also of significance because while the electric field strength is zero at $r=0$, the singularity is actually ``removable'' in the sense that the polarization field can be extended to a continuous function at $r=0$. This then results in the skrymion number being exactly an integer, namely $\pm 1$, respectively. Note however that this does not result in true topological stability as perturbations in coefficients will in general change the asymptotic behaviour of the field near $r=0$ preventing the singularity from being removed. This is also evident from the figure, which shows that perturbations in the coefficients do not preserve the skyrmion number when $B_{11}=0$. 

\begin{figure}[!ht]
\centering
\includegraphics[width=1\textwidth]{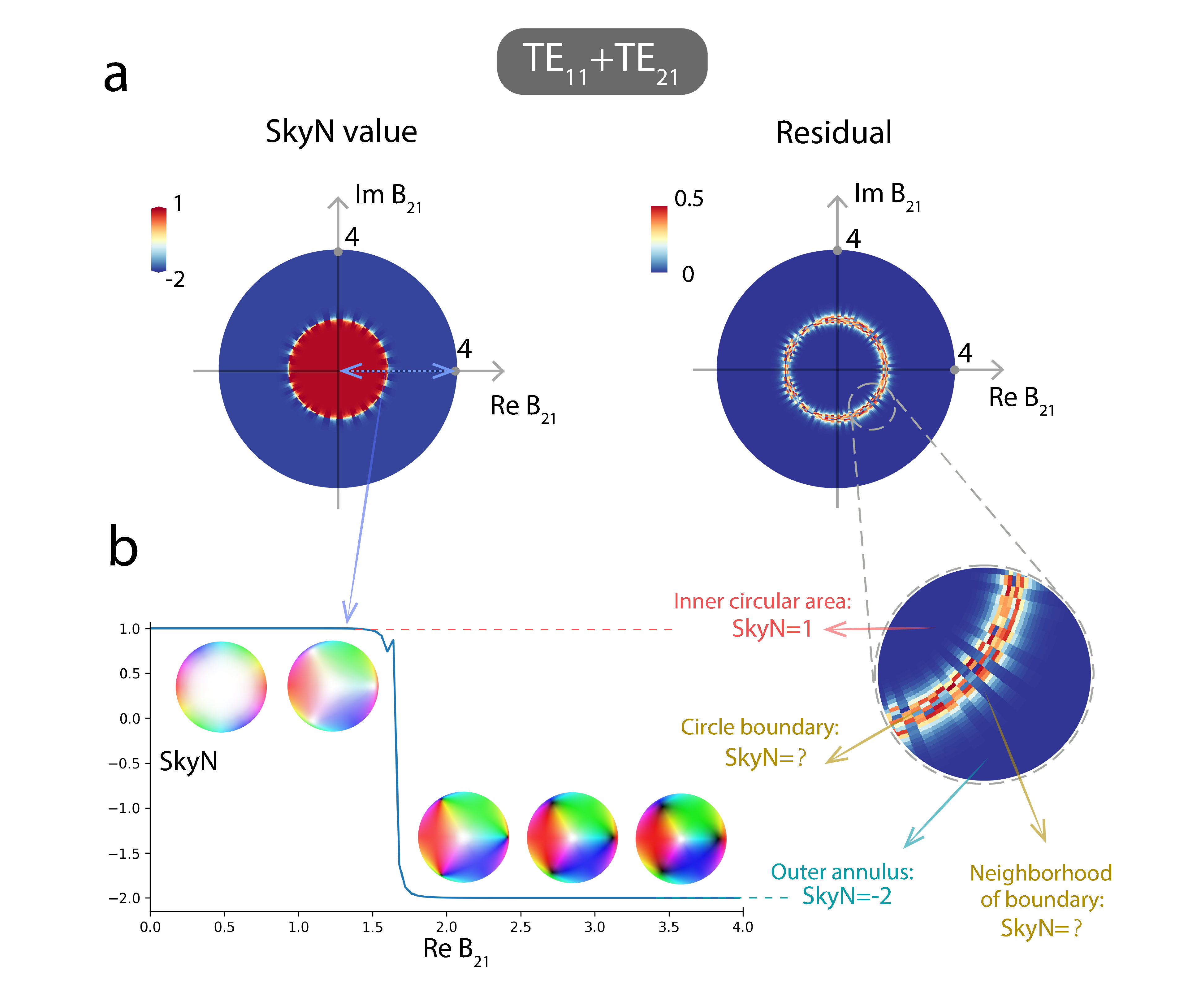}
\caption{\footnotesize \label{fig: TE11+TE21} {\bf Phase diagram of a superposition of TE\textsubscript{11} and TE\textsubscript{21} modes.} a, (Left) Skyrmion number at different values of $B_{21}$. (Right) Absolute difference between computed skyrmion number and its nearest integer. The skyrmion number is stable with the value of $1$ in the inner circular region and stable with a value of $-2$ in the outer annulus. Between the two stable regions, there is a critical region where the behaviour of the skyrmion number is unpredictable, with no intermediate topological phase in between. b, Skyrmion number along the $\text{Im}\, B_{21}=0$ line. The stokes fields at different $\text{Re}\, B_{21}$ taken uniformly between 0 and 4 are also shown.}
\end{figure}

Moving on to the dispersive case, we consider a superposition of $\text{TE}_{11}$ and $\text{TE}_{21}$ modes. First, we fix $z=0$ and study the field for different coefficients. Figure \ref{fig: TE11+TE21} shows the case where $A_{11} = 1$, $B_{11} = 0$, $A_{21} = 0$ and $B_{21}$ is varied. From the figure, we see that, as in figure \ref{fig: TE11}, we once again recover two topologically stable regions, an inner disk where the skyrmion number is stable with the value $1$, and an outer region where the skyrmion number is stable with the value of $-2$. This demonstrates that the $\text{TE}_{21}$ mode has been stabilized by the $\text{TE}_{11}$ mode, with the singularity at $r=0$ no longer present. Importantly, notice that by superposition, we have produced a stable configuration with a skyrmion number of $-2$, which is impossible to achieve with either mode alone. As in Figure \ref{eq: TE1}, there is a critical region between the two stable modes; however, in this case, there is no intermediate topological phase.

\begin{figure}[!ht]
\centering
\includegraphics[width=0.95\textwidth]{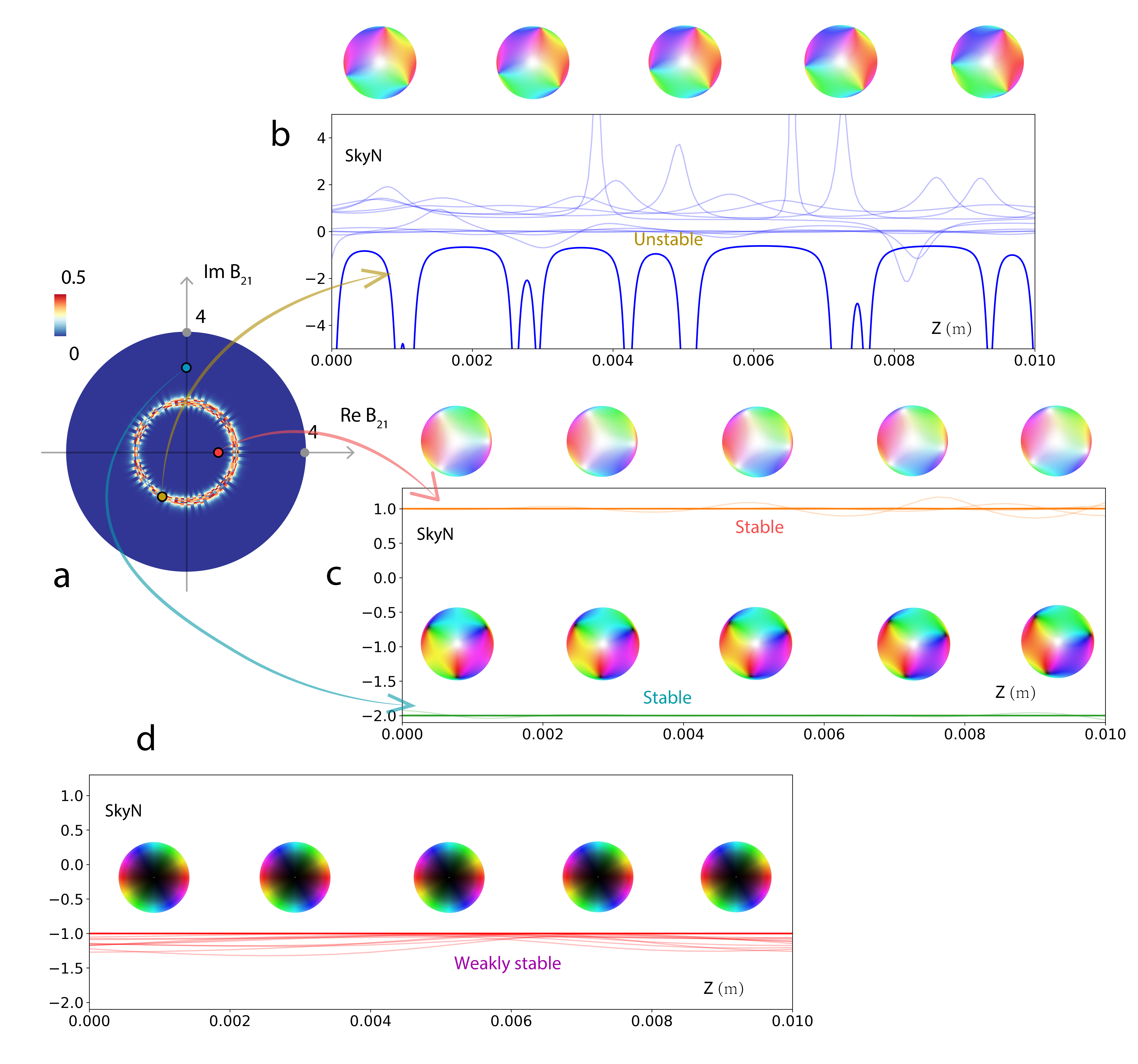}
\caption{\footnotesize \label{fig: propagation} {\bf Skyrmion number of dispersive modes along a waveguide.} a, Chosen stable and unstable modes marked on the relevant phase diagram. b, Skyrmion number and stokes field of an unstable superposition of TE\textsubscript{11} and TE\textsubscript{21} modes ($A_{11}=1$, $B_{21}=0.7-1.5i$) against distance along the waveguide. The solid line shows the skyrmion number of the unperturbed mode, while the translucent lines show the skyrmion number of perturbed modes. c, Skyrmion number and stokes field of two stable superposition of TE\textsubscript{11} and TE\textsubscript{21} modes (orange: $A_{11}=1$, $B_{21}=1$, green: $A_{11}=1$, $B_{21}=2.5i$) against distance along the waveguide. The figure is organized in a similar way to $(a)$. d, Skyrmion number and stokes field of a weakly stable mode ($B_{21}=1$) against distance along the waveguide. The skyrmion number of the unperturbed mode is exactly $-1$, but is not stable to perturbations in coefficients. The figure is organized in a similar way to $(a)$.} 
\end{figure}

More importantly, we demonstrate in figure 3 the skyrmion number of various dispersive situations as $z$ changes. In each case, we consider the superposition of two modes with a given coefficient and stability (as represented by the solid lines), then add random perturbations to the four relevant coefficients uniformly drawn from $[-0.25,0.25]\times [-0.25, 0.25]\subset \mathbb{R}^2 = \mathbb{C}$ (as represented by the translucent lines). The phase diagram in figure \ref{fig: propagation}a shows the position of the respective modes, and for each chosen mode, 10 random samples are drawn.

The orange and green lines (figure \ref{fig: propagation}c) are given by a topologically stabilizing superposition of $\text{TE}_{11}$ and $\text{TE}_{21}$ modes. From the figure, it is clear that there is stability of the skyrmion number in both $z$ and with respect to coefficients, with most random instances remaining topologically stable in propagation with the same topological number. Notice from the phase diagram that since the orange point is closer to the critical region, it is more easily perturbed into an unstable state. This explains the slightly larger fluctuations in the randomly drawn cases corresponding to the orange line as compared to green line. 

The blue line (figure \ref{fig: propagation}a) is given by an unstable superposition of $\text{TE}_{11}$ and $\text{TE}_{21}$ modes. In this case, there is neither stability in $z$, nor stability with respect to coefficients, and perturbations to coefficients result in drastically different skyrmion number profiles along the waveguide.

Lastly, the red line (figure \ref{fig: propagation}d) shows a weakly stable mode, namely $B_{21}=1$ with all other coefficients equal to $0$. In this case, the unperturbed mode is dispersionless so both the stokes field and skyrmion number remains unchanged in propagation. However, changes in coefficients affect the stability of the skyrmion number, as shown in the figure. Here, perturbations are added to the coefficients of $A_{21}$, $B_{21}$, $A_{22}$ and $B_{22}$.

\clearpage

\section{Discussion}

In this paper, we have developed, to the best of our knowledge, the first theory of optical skyrmion transport in waveguides. Our theory classifies solutions to the Helmholtz equation within a waveguide into three distinct categories: stabilizing modes, weakly stable modes, and unstable modes. Stabilizing modes are solutions that are non-zero throughout the waveguide and exhibit both topological robustness in propagation and to changes in coefficients. Moreover, these modes can be used to stabilize other modes. Weakly stable modes are solutions that are topologically protected during propagation but not to perturbations in coefficients. This is indicative of a dispersionless mode with singularities or modes with removable singularities. Lastly, unstable modes are solutions that are neither topologically protected in propagation, nor to perturbations in coefficients. 

The framework introduced above offers a comprehensive solution to skyrmion transport within a conducting waveguide and can be extended to overcome changing boundary conditions and polarization singularities by adopting the generalized skyrmion number introduced in\cite{wang2024generalizedskyrmions}. This is particularly important for two reasons. Firstly, in dielectric waveguides, the polarization state at the boundary is not fixed along the length of the waveguide. Secondly, the ability to utilize modes with singularities significantly expands the range of topological information that can be transported within a waveguide.

The generalized skyrmion number is a way of assigning an arbitrary $S^2$-valued field which cannot be compactified with multiple topological charges depending on the nature of its boundary. In particular, one can assign a distinct topological number to each connected component carved out by the curve traced by boundary polarization states on the Poincar\'{e} sphere. The theory also handles singularities, which behave as extra boundaries. Importantly, the generalized skyrmion number is preserved under a broad class of deformations---as long as the number of connected components carved out by the boundary remains unchanged in propagation, so too does the generalized skyrmion number. Additionally, if a connected component disappears as the field propagates, the topological charges carried by the remaining components remain unchanged. Therefore, the generalized skyrmion number offers topological robustness in many real-world settings.

\begin{figure}[!ht]
\centering
\includegraphics[width=0.85\textwidth]{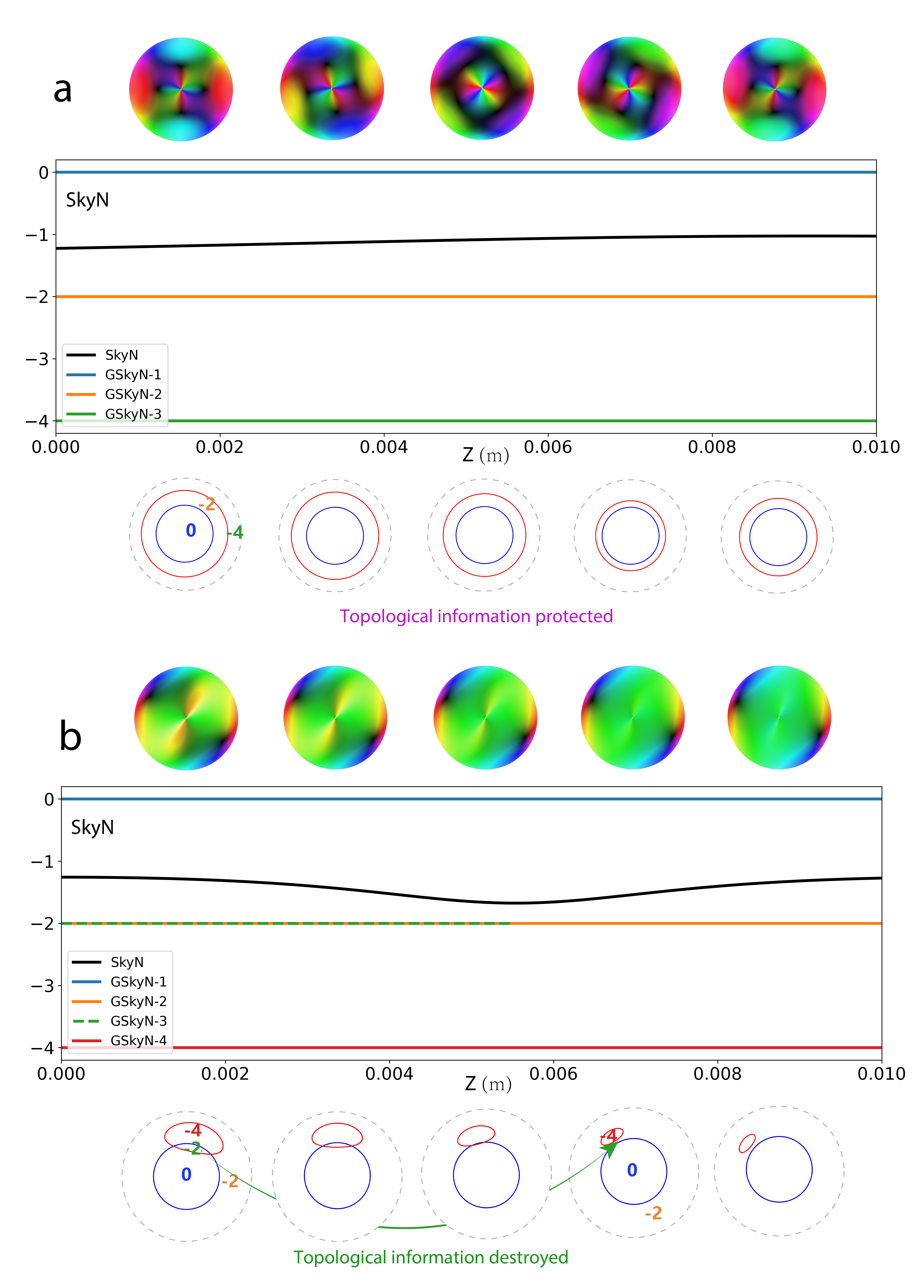}
\caption{\footnotesize \label{fig: generalized} {\bf Generalized skyrmion number in propagation.} a, The skyrmion number and generalized skyrmion number of a superposition of TE\textsubscript{21} and TE\textsubscript{22} modes with coefficients $B_{21}=1$, $A_{22}=0.5$ and $B_{22}=0.3$ at different $z$. The stokes fields within the waveguide at different $z$ drawn uniformly between $0$m and $0.01$m are shown above the plot, and the stereographically projected image of the waveguide's boundary (blue) and singularity's boundary (red) at different $z$ drawn uniformly between $0$m and $0.01$m are shown below the plot. The generalized skyrmion numbers corresponding to each component is also shown. Since the number of components remains unchanged in propagation, topological information is protected. b, The skyrmion number and generalized skyrmion number of a superposition of TE\textsubscript{01} and TE\textsubscript{21} modes with coefficients $A_{01}=-0.4i$, $B_{01}=0.15$, $A_{21}=0.15i$ and $B_{21}=0.7$. The figure is organized in a similar way to (a). Here, the component corresponding to third generalized skyrmion number disappears in propagation, and the topological charge of that component is lost. However, the remaining generalized skyrmion numbers are preserved in propagation. }
\end{figure}

To demonstrate the power of the generalized skyrmion number, we present in figure \ref{fig: generalized} two scenarios in a cylindrical conducting waveguide where the usual skyrmion number is not preserved. Figure \ref{fig: generalized}a is a superposition of TE\textsubscript{21} and TE\textsubscript{22} modes with coefficients $B_{21}=1$, $A_{22}=0.5$ and $B_{22}=0.3$. In this case, there is a persistent singularity at $r=0$, as can be seen by the stokes fields in figure \ref{fig: generalized}a, which causes the usual skyrmion number to fluctuate between $-1.23$ and $-1.03$ in propagation. On the other hand, the boundary curve and singularity carve out three distinct regions on the Poincar\'e sphere for all $z$, which give rise to three generalized skyrmion numbers. The evolution of these curves in propagation is represented by the red and blue circles in figure \ref{fig: generalized}a below the plot of skyrmion numbers. The blue curve is the image of the polarization states along the physical boundary of the waveguide on the Poincar\'{e} sphere after stereographic projection from the north pole, while the red curve is the image of the boundary curve associated with the singularity. From the figure, it is clear that the number of components that the Poincar\'{e} sphere is partitioned into remains unchanged in propagation. As a result, the generalized skyrmion number remains stable at $(0, -2, -4)$ throughout. 

Figure \ref{fig: generalized}b shows a superposition of TE\textsubscript{01} and TE\textsubscript{21} modes with coefficients $A_{01}=-0.4i$, $B_{01}=0.15$, $A_{21}=0.15i$ and $B_{21}=0.7$. As above, there is a persistent singularity at $r=0$ which cases the usual skyrmion number to fluctuate between $-1.26$ and $-1.68$. From the figure, the number of components that the Poincar\'{e} sphere is partitioned into starts at 4, but transitions to 3 during propagation, and there is a transition of generalized skyrmion number from $(0,-2,-2,-4)$ to $(0, -2, -4)$. Notice, however, that while the topological information of the disappearing component is lost, the topological numbers of the remaining components remain preserved during propagation.

In the two scenarios presented above, not only does the generalized skyrmion number solve the problem of an unstable topological charge, but also significantly enhances information density by enabling the transmission of multiple topologically protected quantities within a single field. 

From a practical perspective, the generalized skyrmion enables full utilization of modes within a waveguide, leveraging multi-mode waveguides in a completely novel way. Moreover, we emphasize that while we only consider conducting waveguides here, a similar methodology can be applied to dielectric waveguides. In the context of dielectric waveguides, the polarization state at the boundary is no longer fixed, and therefore the conventional skyrmion number integral is not expected to be preserved. The generalized skyrmion number, however, remains applicable. More specifically, recall that the topological protection of the generalized skyrmion number is guaranteed as long as the number of components carved out by the boundary remains unchanged during propagation. This is a relatively weak condition which, in the absence of polarization singularities, is certainly satisfied over a finite distance due to the field's continuity, and is expected to hold at length scales relevant to waveguides in integrated photonic circuits. Thus, the approach introduced in our paper is highly promising for practical implementations in photonic device engineering. We leave a full investigation into the propagation of skyrmions in dielectric waveguides to a future paper.

To conclude, this paper discusses the transport of skyrmions in conducting waveguides and demonstrates the stability of both the conventional skyrmion number integral and the generalized skyrmion number within these waveguides. We once again emphasize the importance of waveguides in enabling high-bandwidth, low-latency data manipulation in photonic systems. The possibility of increasing the dimensionality of information transmitted through a waveguide by exploiting polarization is an exciting prospect, and using topology to provide robustness and overcome modal dispersion is, to the best of our knowledge, a completely novel approach to leveraging polarization in such systems. This is particularly true because polarization, an independent degree of freedom alongside phase, amplitude, and wavelength, can be simultaneously modulated both in space and time. Our work not only opens new avenues for robust and efficient high-dimensional information encoding in integrated photonics, but also highlights the broader potential of topological methods in high density data applications such as next-generation optical communications and photonic computing.

\clearpage

\section*{Methods}
\setcounter{section}{0}
\section{TE and TM modes in a cylindrical conducting waveguide}

By simple separation of variables, the TE modes within a cylindrical conducting waveguide of radius $a$ extending in the $z$ direction are given by 
\begin{align}
    E_r^{\text{TE}_{mn}}(r,\theta) & = \frac{i}{(\omega/c)^2-k_n^2}\left(\frac{\omega}{r}\left(imA_{mn}e^{im\theta}-im B_{mn}e^{-im\theta}\right) J_m\left( \frac{rp_{mn}'}{a}\right)\right) \label{eq: TE1} \\
    E_\theta^{\text{TE}_{mn}}(r,\theta) & = \frac{i}{(\omega/c)^2-k_n^2} \times\left(-\omega\left(A_{mn}e^{im\theta} + B_{mn}e^{-im\theta}\right)\frac{p_{mn}'}{a}J_m'\left( \frac{rp_{mn}'}{a}\right)\right) \label{eq: TE2}
\end{align}
where $m\in \mathbb{N}$, $n \in \mathbb{N}^+$, $J_m$ is the order $m$ Bessel function of the first kind, $p_{mn}'$ the $n$\textsuperscript{th} root of $J_m'$ greater than $0$ and 
\begin{equation}
    k_n^2 = \left(\frac{\omega}{c}\right)^2 - \left(\frac{p_{mn}'}{a}\right)^2,
\end{equation}
where the mode is propagating if $k_n^2 > 0$ and evanescent otherwise. Likewise, we have the TM modes
\begin{align}
    E_r^{\text{TM}_{mn}}(r,\theta) & = \frac{i}{(\omega/c)^2-\kappa_n^2}\left(\kappa_n\left(C_{mn}e^{im\theta} + D_{mn}e^{-im\theta}\right) \left(\frac{p_{mn}}{a}\right)J_m'\left(\frac{rp_{mn}}{a}\right)\right) \\
    E_\theta^{\text{TM}_{mn}}(r,\theta) & = \frac{i}{(\omega/c)^2-\kappa_n^2}\left(\frac{\kappa_n}{r}\left(imC_{mn}e^{im\theta}-imD_{mn}e^{-im\theta}\right)J_m\left(\frac{rp_{mn}}{a}\right)\right)
\end{align}
where $m \in \mathbb{N}, n\in \mathbb{N}^+$, $p_{mn}$ is the $n$\textsuperscript{th} root of $J_m$ greater than 0 and 
\begin{equation}
    \kappa_n^2 = \left(\frac{\omega}{c}\right)^2 - \left(\frac{p_{mn}}{a}\right)^2.
\end{equation}

\bibliographystyle{unsrt}
\bibliography{main}

\end{document}